# Does Interdisciplinary Creative Coding Boost Creativity? A Mixed Methods Approach (PRACTICE)

**A. Duyver**
KU Leuven
Diepenbeek, Belgium
0009-0007-1447-1611

**W. Groeneveld**
KU Leuven
Diepenbeek, Belgium
0000-0001-5099-7177

**K. Aerts** [1]
KU Leuven
Diepenbeek, Belgium
0000-0001-7214-452X

## ABSTRACT

This study explores the influence of an interdisciplinary intervention on creative problem-solving skills. Literature deems such skills as vital for software engineering (SE) students in higher education. 39 SE students and graphic design (GD) students were randomly paired to work on an open-ended creative coding assignment in p5.js, an online JS-based Processing editor that makes it easy for novices to quickly and easily code visual webpages. Three categories were formed: the test group SE+GD (18 students), and control groups SE+SE (10) and GD+GD (11).

---

[1] *Corresponding Author, K. Aerts,*

kris.aerts@kuleuven.be

A mixed methods approach was taken to gather and interpret results: Amabile's Consensual Assessment Technique provided a global creativity score for the finished product, the Creative Programming Problem Solving Test assessed three dimensions of the creative process (*Ability, Mindset, Interaction)*, and 9 semi-structured follow-up interviews provided context and revealed underlying themes. The results indicate that, while the creativity of the end product initially takes a hit, the SE+GD groups' socio-interactive creativity levels increased. We also observed fixed mindsets towards creativity ("design students are more creative than we") that call for future work.

## 1 INTRODUCTION

Many studies emphasize creativity as an essential problem-solving skill in the world of computing (Apiola and Sutinen 2020; Salgian et al. 2013). A recent Delphi study reveals: SE industry experts rate creativity as crucial to succeed as a developer (Groeneveld et al. 2020). Another study concluded that in order to foster creativity in higher education, three approaches can be taken (Groeneveld, Becker, and Vennekens 2021):

- Introduce experimental learning. Experiments with open-ended project-based learning have been shown to be beneficial towards students' creativity.
- Get students out of the classroom. The environment in which the learning takes place plays a significant part in the creative potential of students.
- Put creativity first, programming second. This opens up computing to a more diverse student population.

This paper provides an experience report that combines all three approaches and adds upon it by introducing an interdisciplinary approach for developing SE student's creative skills. In our approach SE students are paired with GD students in an experimental open-ended learning project using creative coding in the programming framework p5.js. The project took place in the design classrooms of another university. And lastly, we put creativity first by refraining from grading and emphasizing on experimentation and having fun as a goal in itself.

The goal was to amplify creative problem-solving capabilities of SE students in higher education, beyond the conventional approaches. Since creativity can express itself in different ways, we ask the following questions:

- **RQ1**: *What is the influence of an interdisciplinary creative coding project on the creativity of the **process**?*
- **RQ2**: *What is the influence of an interdisciplinary creative coding project on the creativity of the **end product**?*

The remainder of this paper is structured as follows. Section 2 outlines related work, Section 3 describes the mixed methods approach utilized to gather results. Those results are presented and discussed at length in Section 4. Next, in Section 5, we highlight possible limitations of this work. Finally, Section 6 concludes this research and suggests future work.

## 2 BACKGROUND

### 2.1 Interdisciplinary collaboration

In the paper "Ten Cheers for Interdisciplinarity" (Nissani 1997) creativity is considered one of the reasons to pursue interdisciplinary collaborations. "Interdisciplinary computing classes are worth the effort", concludes Lori Carter (Carter 2014). We believe this to be especially relevant in the field of computing where the demand for interdisciplinary skill crossovers is growing, according to (Carr, Jones, and Wei 2020).

### 2.2 Creative coding

The concept "creative coding" is often employed in curricula to explore code as a medium for self-expression (Peppler and Kafai 2009). The focus here is not sparking creativity to deal with daily programming problems, but rather to use code to express your creative urge. This is typically done using the Processing (p5.js) programming language and has been known to increase engagement and excitement for computing (Greenberg, Xu, and Kumar 2013). The focus on visual creations entices students, while at the same time offering a decent programming challenge.

### 2.3 Creativity and how to assess it

Creativity is a broad concept that, even when viewed from a computational perspective, seems to codify multiple perspectives: creativity by yourself, in teams, or on socio-organizational levels (Veale, Gervás, and Pease 2006). Precisely because of ongoing discussions whether or not the concept of creativity is context-dependent (Baer 2010) and disagreements on numerous definitions of creativity (Groeneveld, Becker, and Vennekens 2021), a plethora of assessment techniques have been published. Yet many existing assessment tools, including those from the field of cognitive psychology, fall short of measuring multiple dimensions of creativity. For instance, the well-known Torrance Test of Creative Thinking (Torrance 1972) gauges divergent thinking but ignores creative collaborative aspects. Personality-based self-tests such as The Big Five emphasize individual creativity and motivation (Sung and Choi 2009). Amabile's Consensual Assessment Technique (CAT) employs a jury to score the creative output but ignores the creative process (Amabile 1982). Recently, a new creativity self-assessment tool was developed specifically geared towards problem solving for computing students, called the Creative Programming Problem Solving Test (CPPST) (Groeneveld et al. 2022). It contains three overarching constructs of creativity based on existing validated scales and conducted focus groups: *Ability* (knowledge of coding and creative techniques), *Mindset* (curiosity and belief in own abilities), and *Interaction* (social aspects of creativity).

## 3 METHODOLOGY

Before elaborating on the data gathering processes, to help the reader interpret the results, we first describe the target groups involved in this study.

## 3.1 The Setting

Students from two entirely different programs took part in the experiment: 19 second-year students from our local faculty of engineering technology electronics/ICT and informatics (SE) and 20 students from the faculty of visual design from a neighboring university (GD). As creative skills are a learning outcome for both groups of students, all students were expected to participate, but to avoid pressure no grades were attached to the process nor to the results. Students could also bail out of filling out any form or participating in the interviews without any consequences, resulting in a setup following the guidelines of the Privacy and Ethics Unit of the university.

All participants were randomly placed into one of three groups: The test group SE+GD (18 students, 9 duos), the control group SE+SE (10 students, 5 duos) or the control group GD+GD (11 students, 4 duos and 1 trio). None of the students had prior experience with the p5.js framework. So an introduction session of two hours was given to all participants. In this study, the decision to use Processing was not made to explore creativity as a means for self-expression like in (Peppler and Kafai 2009), but as a means for problem solving. To minimize unwanted side effects during measurements, all groups were placed in the same physical location, including control groups SE+SE and GD+GD. We chose the buildings of the design faculty for the location since we were most interested in the possible deltas of SE students and, as explained before, wanted to "get students out of the classroom". Of course, for the design students, the location didn't change. As for the project assignment itself, it was delineated, but not too much, as to leave room for creative freedom. The pairs had to create a visual exposition that emphasizes user interaction, for instance through sound, camera, or mouse input. The assignment had to be completed on a single day. All projects were incorporated into an online p5.js exposition.

## 3.2 Measuring The End Result: CAT

Well beyond the field of cognitive psychology, Amabile's CAT is commonly used to evaluate the creativity of an end product (Baer and McKool 2009), in our case, each p5.js project. CAT relies on expert judges that score the creativity of an end product between 1 and 10. Since the scoring process is very subjective, it is recommended to recruit multiple judges and work with an average. For this study, we enlisted seven judges that are part of the teaching staff of the involved courses: three computing experts (all co-authors), and four GD experts, of which two eventually opted out of the study. The judges were instructed in individually evaluating the creativity of each anonymized project by spending exactly one minute on each project. We refrained from providing a definition of creativity, as per recommendation in (Baer and McKool 2009). The standard deviation ($SD$) of the scores, on average 1.44, mirrored Baer and McKool's conclusion: judges score surprisingly similar (Baer and McKool 2009).Therefore, for CAT, inter-rater reliabilities are not relevant. However, as an extra verification step, when judges did not agree (threshold of $SD > 1.80$, 5 out of 19 or 26% of the projects), those projects were re-discussed in group, after which new scores were assigned and an $SD$ calculated, until the threshold was reached.

### 3.3 Measuring The End Result: CPPST

Next to evaluating the end result with CAT, we were also interested in different aspects of the creative process. The CPPST tool allows us to assess whether or not our interdisciplinary intervention has effect on specific parts of students' creative problem solving abilities. The CPPST is a self-assessment test which was administered at the end of the project day. In it, students answer 56 questions on a Likert-5 scale. The full question set is available in (Groeneveld et al. 2022). A reduced set of 32 questions are enough to gauge the three factors of the CPPST, but we opted to include the full set as the extra data might help us in asking more specific questions in the semi-structured interviews.

### 3.4 Enriching quantitative data: interviews

The different CAT and CPPST values are devoid of rich contextual information. Therefore, we decided to conduct additional interviews to put these numbers into context based on the results of the aforementioned tests. After the CPPST survey results were collected, average and $SD$ values were calculated for each question, grouped by the three student categories. Next, since we were most interested in the impact of the interdisciplinary component, deltas ($\delta$) of averages between the subgroup SE from SE+GD vs SE+SE and the subgroup GD from SE+GD vs GD+GD were calculated. A $\delta > 0.5$ was marked as potentially interesting. Hove and Anda's recommendations for conducting semi-structured interviews in empirical SE research (Hove and Anda 2005) combined with a discussion of the deviating $\delta$ values of the CPPST results guided us to the following question sets, divided into two distinct themes: **general context** (1-2) and more **in-depth**-related (3-9) questions:

1) What did you think of the experience?
2) How did the collaboration go?
3) How did you tackle the project in general?
4) How did you tackle brainstorming and ideation?
5) What did you do when a problem occurred and you were stuck?
6) What did you learn with this project?
7) In which way did you get out of your comfort zone?
8) How did you tackle the openness of the assignment?
9) What would you do different if you were to re-do it?

Our aim was to interview a random selection of 30% of the participants of each group. Data from the interviews was processed using qualitative coding as presented by Onwuegbuzie et al. (Onwuegbuzie et al. 2009). The transcripts were read multiple times independently by two co-authors to apply an open coding step, initially identifying 43 codes. Next, in order to identify patterns, notes were compared, cross-validated, and reduced into 25 codes grouped into 4 themes in an axial coding step. These themes served as a starting point for discussion and to cross-link back to the quantitative results. The resulting themes and codes are presented in section 4.

## 4 RESULTS AND DISCUSSION

At the end of the project day, all pairs successfully created an interactive visual exposition. An overview of the projects, together with all open data used in this study, is available at https://arneduyver.github.io/creative-coding/gallery. Some examples:

- Abstract art that reacts to sound or mouse input.
- "Draw in the air" using the camera.
- A text to music generator.

### 4.1 Quantitative results

The CAT scores of the SE+GD, SE+SE and GD+GD groups were respectively 6.30, 6.88 and 6.04. Although these differences can be considered small, it is interesting to note that the test group of interdisciplinary teams (SE+GD) scored with 6.30 less than the pure SE teams (6.88). To verify if the differences originate from the creative process, we took a closer look at the CPPST measurements (on a scale of 5).

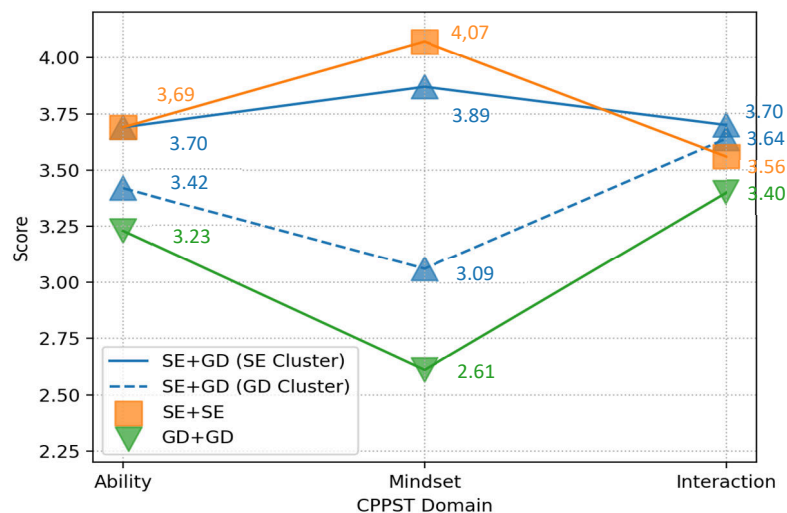

Figure 1: Average CPPST domains for groups SE+GD (△), SE+SE (□), GD+GD (▽).

Figure 1 shows that the interdisciplinary test group performed marginally better at *Interaction* level, and especially in the socio-interactive components of creativity, probably because their different backgrounds enforce more discussion. However, the biggest gap is in *Mindset* with low scores for for the GD+GD group, but also for the SE+GD groups with a GD majority. It might be that they struggled with the complexity of programming and had no access to a SE student to solve this, or it might be a sign of a fixed mindset ("I can't code") that resulted in an early defeat. As CAT and CPPST are devoid of context, a qualitative interpretation is needed.

### 4.2 Qualitative results

In total, 9 students were interviewed: 5 from the SE+GD group (3 from SE and 2 from GD, thus 29.4% of that population), 3 from the SE+SE group (33.3%), and 1 from the GD+GD group (9%). Since we are mostly interested in the effects of the intervention on the SE students, we do not consider the small GD sample size as a threat to the validity of the study. Transcript analysis initially yielded 43 codes across all groups, reduced to 25 and categorized in 4 distinct themes: *Curiosity* (6), *Cooperation* (7),

*Method* (7), and *Mindset* (5). We now briefly describe findings of each of the themes with codes emphasized in bold.

**_CURIOSITY_** - Almost all interviewees appreciated the opportunity given to **explore beyond their education**. SE students mentioned they "*had no idea it was that easy to create something visually*". Especially **being in the same space** turned out to be inspirational, as students regularly got up to see others at work and mentioned ideas cross-pollinated quicker that way. The unique **physical environment**, as per recommendation in (Groeneveld, Becker, and Vennekens 2021), seemed to inspire SE students:

> *I was very impressed because as you entered, there were drawing tablets, 3D printers, a green room for photography, ...Then more ideas will come to you.*

Another trigger for curiosity was p5.js itself, as the **framework facilitated play**: students praised the clarity of the documentation and the easy-to-use web editor that facilitates experimentation through rapid feedback. The **freedom** and absence of the stress in anticipation of a grade was also reported to play a role. The interviews also revealed that **frustration** was a big factor limiting creativity.

**_COOPERATION_** - SE+SE students working in a homogeneous group mentioned cooperation was **smooth**, even if they did not know their partner. Yet, for the heterogeneous SE+GD group, collaborating was **very difficult**. "*Working together with a complete stranger was hard.*" When asked why, several reasons were given: (1) It was "*difficult to explain your own way of thinking*"; (2) "*I really had to drag it all out [of my partner]*". (3) Some students wanted to get to know their partner, while others wanted to start coding immediately. The different fields of study might also imply a difference in personality and approach. **Giving feedback** proved to be a challenge as well. The interviews confirm that the social aspects of co-creating was what pushed students **out of their comfort zone**. Some homogeneous interviewees realized that cooperating--and possibly, the creative outcome--would be **very different in a mixed group**, but in a good way: "*I found projects from the mixed group to be excelling [...] they had ideas that were more interesting than ours*". Although "*more interesting*" did not result in higher average CAT scores, all students do **acknowledge that cooperation is an important factor** to creative success.

**_METHOD_** - SE+SE and GD+GD groups applied **different approaches**. The former look at examples, think about what they like to do based on the examples, and immediately start exploring that in code, while the latter first brainstorm for ideas, sometimes **iterating** over them, and only then look at what is possible to try and implement their ideas. SE students mentioned they had **difficulties with the openness of the assignment**. Some SE+SE students admitted spending too much time trying to come up with a concept, struggling with what they wanted to create. For SE+GD groups, there was a **stereotypical task division**: "you *do the code, I'll do the design*". As to what to do when stuck, SE+GD students mentioned they liked to "***get up and walk around** [...] I take that as a moment to think about something else [...] and usually come back with a solution*". SE+SE students prefer **diving into the docs** before asking for help.

***MINDSET*** - Interviews revealed many **prejudices**: from "*creating something visual is hard*" (SE) to "*I can't code*" (GD). As much as everyone acknowledged **importance of heterogeneity in teams** and lauded the experiment as **refreshing**, we discovered a **fixed mindset** when it comes to creativity. SE students say "*design students go more in the creative direction while we use logical steps to solve things*", while GD students say "*coding is very mathematical*" and imply it is less creative than their visual work. Clearly, this fixed mindset can be very damaging to the creative potential of SE students. Also, students have a **wrong image** of what an engineer or designer does, as a GD student testified:

> *I think we're more the people who ask questions while you just tell an engineer to do this or that and he'll understand it that way.*

## 5   LIMITATIONS

This paper reports on an intervention with a relatively small group ($n = 39$). Although judging from the recurring themes throughout the interviews, we believe that the findings of this paper will persist. Since the cooperation between different student groups caused friction, a prolonged intervention of for instance a week could iron out the initial acquaintance difficulties. We suspect that the CPPST *Interaction* $\delta$ between the groups would even be bigger then, possibly also increasing *Ability*, but the creativity prejudices would likely remain. Future work might shed more light on this.

## 6   CONCLUSION

This study explored the influence of an interdisciplinary intervention on creative problem-solving skills by pairing up SE students with GD students. A mixed methods approach helped in identifying and understanding the various effects of the intervention. While we observed a slight decrease in the creativity of the end product (RQ1), the CPPST reveals that although *Mindset* needs more work, our intervention effectively increased the *Interaction* part of the creative problem-solving process (RQ2). It is important to note that due to the various reasons touched upon while discussing the qualitative results in Section 4.2, creativity can initially take a hit during interventions. This should not worry educators but is something to be aware of. Using a measurement such as CAT is not enough to reveal the underlying constructs behind the numbers. As many students testified, the intervention was a great experience to get a taste of *real* creative cooperation, even though this is not visible by looking at the CAT results. And yet, a recent literature review revealed that most computing education studies on creativity employ a single metric (Groeneveld, Becker, and Vennekens 2021). We also found the fixed mindset approach towards creativity to be very problematic, especially for the teaching staff who try to improve the creative skills of students. This has also been noted by Apiola & Sutinen (Apiola and Sutinen 2020) and Groeneveld et al. (Groeneveld et al. 2022). We have not encountered any studies that try to mitigate this. We see no easy solution and thus feel that this requires immediate attention from the computing education community.

**ACKNOWLEDGEMENTS**

Thanks to all students and creativity judges for their enthusiasm and interest.